\documentclass[bibyear]{aa}

\usepackage{amsmath}
\usepackage[utf8]{inputenc}
\usepackage[T1]{fontenc}
\usepackage[english]{babel}
\usepackage{graphicx}
\usepackage{xcolor}
\definecolor{grey93}{rgb}{0.93,0.93,0.93}
\usepackage{color}
\usepackage{colortbl}
\usepackage{xtab}
\usepackage{longtable}
\usepackage{tabularx}
\usepackage{url}
\usepackage{caption}
\usepackage{amssymb}
\usepackage[dvipdfm]{hyperref}
\usepackage{fancyvrb}
\usepackage{multirow}
\usepackage[varg]{txfonts}
\usepackage{natbib}
\usepackage{lscape}

\hypersetup{
	pdfpagemode=UseOutlines,      
	pdfstartview=Fit,             
	pdffitwindow=true,            
	pdfpagelayout=TwoColumnsRight,
	pdftoolbar=true,              
	pdfmenubar=true,              
	bookmarksopen=false,          
	bookmarksnumbered=true,       
	colorlinks=true,              
	pdfauthor={Ton nom},          
	pdftitle={Titre PDF},         
	pdfcreator=PDFLaTeX,          %
	pdfproducer=PDFLaTeX,         %
	linkcolor=blue,               
	urlcolor=blue,                
	anchorcolor=black,            
	citecolor=blue,               
	frenchlinks=true,             
	pdfborder={0 0 0}             
}

\begin{document}

\title{Cassini ISS Mutual Event Astrometry of the Mid-sized Saturnian Satellites 2005-2012\thanks{Full Table 4 and Table 5 are only available at the CDS via anonymous ftp to \href{cdsarc.u-strasbg.fr (130.79.128.5)}{cdsarc.u-strasbg.fr (130.79.128.5)} or via \href{http://cdsweb.u-strasbg.fr/cgi-bin/qcat?J/A+A/.}{http://cdsweb.u-strasbg.fr/cgi-bin/qcat?J/A+A/.}}}

\author{N.J. Cooper\inst{\ref{qmul}, \ref{imcce}}
\and C.D. Murray\inst{\ref{qmul}}
\and V. Lainey\inst{\ref{imcce}}
\and R. Tajeddine\inst{\ref{Cornell}, \ref{imcce}}
\and M.W. Evans\inst{\ref{Cornell}}
\and G.A. Williams\inst{\ref{qmul}}}

\institute{Astronomy Unit, School of Physics and Astronomy, Queen Mary University of London, Mile End Road, London, E1 4NS, UK \label{qmul}
\and IMCCE, Observatoire de Paris, UMR 8028 du CNRS, UPMC, Universit\'e de Lille 1, 77 av. Denfert-Rochereau, 75014 Paris, France\label{imcce}
\and Department of Astronomy, Cornell University, Ithaca, NY 14853, USA \label{Cornell} }
\date{Received 8 July 2014 / Accepted  1 September 2014}

\abstract
{}
{We present astrometric observations of the Saturnian satellites Mimas, Enceladus, Tethys, Dione and Rhea from $Cassini$ Imaging Science Subsystem (ISS) narrow-angle camera (NAC) images. Image sequences were designed to observe mutual occultations between these satellites.}
{The positions of satellite centres were estimated by fitting ellipsoidal shape models to the measured limbs of the imaged satellites. Spacecraft pointing corrections were computed using the UCAC2 star catalogue. We compare observed-minus-computed (O$-$C) residuals based on inter-satellite separations with those based on individual satellite positions, relative to the SAT360 and NOE-6-2012-MAIN ephemerides.}
{We provide a total of 2303 astrometric observations, resulting in 976 pairs, the remainder consisting of observations of a single satellite. We obtain mean residuals for the individual satellite positions relative to the SAT360 ephemeris of 4.3 km in the line direction and -2.4 km in the sample direction, with standard deviations of 5.6 and 7.0 km respectively, an order of magnitude improvement in precision compared to published HST observations. We show that, by considering inter-satellite separations, uncertainties in camera pointing and spacecraft positioning along with possible biases in the individual positions of the satellites can be largely eliminated, resulting in an order-of-magnitude increase in accuracy compared to that achievable using the individual satellite positions themselves. We demonstrate how factors relating to the viewing geometry cause small biases in the individual positions of order 0.28 pixel to become systematic across the dataset as a whole and discuss options for reducing their effects. The reduced astrometric data are provided in the form of individual positions for each satellite, together with the measured positions of reference stars, in order to allow more flexibility in the processing of the observations, taking into account possible future advances in limb-fitting techniques as well as the future availability of more accurate star catalogues, such as those from the GAIA mission.}
{}

\keywords{astrometry, occultations, planets and satellites:general, methods:observational}

\maketitle

\section{Introduction}

A planned campaign of astrometric observation of the inner satellites of Saturn using the Imaging Science Subsystem (ISS) of the $Cassini$ orbiter has been ongoing since Saturn Orbit Insertion (SOI) in July 2004. This work has been driven both by scientific objectives and as a contribution to the operational navigation effort of the Cassini project, with regular deliveries of observations provided to JPL for the updating of satellite orbit models throughout the mission. Recent scientific results, see for example \citet{Lainey12}, have highlighted the key role of high-resolution imaging and astrometry techniques in the solution of fundamental problems relating to the structure and dynamical evolution of planetary satellite systems. The data presented here represent a further contribution towards that wider goal.

In terms of previously published Cassini ISS astrometry, observations of the Jovian satellites Amalthea and Thebe using images from Cassini's Jupiter fly-by were published by \citet{Coop06}, while \citet{Taj13} published astrometry of the Saturnian satellites Mimas and Enceladus, using a variety of ISS images, including some from the planned programme described above. \citet{Coop14} present astrometry of the small inner satellites of Saturn, Atlas, Prometheus, Pandora, Janus and Epimetheus using images both from the planned programme and image sequences designed to study Saturn's F ring.

The planned campaign of astrometric data collection has been divided into two parallel programmes: image sequences targeting specific individual satellites, normally identified by the label SATELLORB in the Cassini image sequence name, and sequences designed to capture chance occurrences of more than one satellite in the NAC field-of-view, typically with sequence name containing MUTUALEVE. The observations presented in this paper form part of the second programme.

Although we describe these observations as mutual events, we use the term somewhat more loosely to describe any partial-occultation of one satellite with another, or the occurrence of more than one satellite (without actual or partial occultation) within an image. Since all the target satellites in these images are fully resolved, we have reduced the observations using an astrometric approach based on limb-fitting, rather than the traditional photometric approach adopted for ground-based mutual phenomena for unresolved satellites, based on light curves \citep{Thuillot01}. We described our approach in the next section. 

During the course of this work, we have also investigated potential sources of bias in the astrometric observations and discuss both their origin and possible approaches to reducing their effects.

Throughout, we use the Cassini ISS convention of referring to the image pixel coordinate along the $x$ axis as `sample' and the $y$ coordinate as `line'.

\section{Observations}
\noindent
The images in each sequence used in this work were designed to target a `primary' satellite while taking a series of images as the secondary satellite moved across the field-of-view. A selection of images from a typical sequence (ISS\_144RH\_MUTUALEVE002\_PRIME) is provided in Fig.1, showing the secondary satellite, in this case Dione, entering the field-of-view from the bottom of the first image (a) before being fully occulted by the primary satellite, Rhea, which remains fixed in the centre of the images. Dione then emerges from behind Rhea as the sequence progresses. The images are consecutive and approximately 34 seconds apart, except that two additional images in the sequence, in which Dione is still fully occulted, are not shown. 

\begin{figure}
\resizebox{\hsize}{!}{\includegraphics{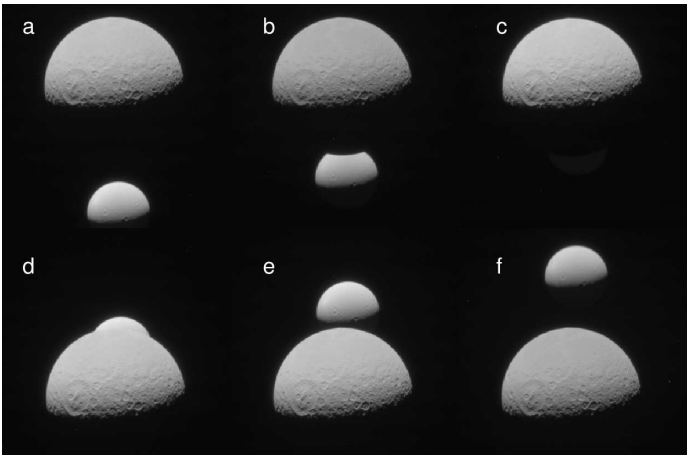}}
\label{Fig. 1}
\caption{A portion of selected images from sequence ISS\_144RH\_MUTUALEVE002\_PRIME, showing Dione being occulted by Rhea. Images are (a) N1675397022\_1.IMG (b) N1675397057\_1.IMG (c) N1675397090\_1.IMG (d) N1675397192\_1.IMG (e) N1675397226\_1.IMG and (f) N1675397261\_1.IMG. The images are consecutive and taken approximately 34 seconds apart, except that images N1675397124\_1.IMG and N1675397159\_1.IMG, between (c) and (d) are not shown, since Dione is fully occulted for these images, as in image (c). Image exposure lengths are (a) 560 ms (b) 2000 ms (c) 460 ms (d) 460 ms (e) 560 ms (f) 2000 ms.}
\end{figure}

Table 1 summarises some relevant characteristics of all the image sequences used in this work. Exposure lengths varied from 60 to 2000 ms. Solar phase angles (observer-object-Sun) varied from 32.5 to 162.9 degrees, with a mean of 103.1$\pm$30.5 deg, while image resolution for the primary satellite ranged from 5.2 to 25.8 km/pixel, with a mean of 13.6$\pm$4.8 km/pixel. Images at the start and end of each sequence generally only contain the primary, hence there are typically more observations of each primary satellite than its corresponding secondary (Table 1). Thus, while the total number of individual observations is 2303, there are 976 pairs of observations (85$\%$ of the total) with the remaining 15$\%$ consisting of observations from images with only one satellite present. Image size is, in all cases, 1024 by 1024 pixels.

\begin{table*}
\caption{Image Sequences}
\label{table:1}
\resizebox{\textwidth}{!} {
\begin{tabular}{llcrrrrrr}
\hline\hline
{Sequence} &{Time (UTC) \tablefootmark{a}}& {Exposure} & {Mean phase\tablefootmark{b}} & {Mean resolution\tablefootmark{b}} & {Primary} & {No.\tablefootmark{c}} & {Secondary}&{No.\tablefootmark{c}}\\
{}&{}&{(ms)} &{(deg)}&{(km/pixel)}&{satellite}&&{satellite}&\\
\hline
ISS\_003DI\_MUTUALEVE004\_PRIME &2005 FEB 20 12:26:06.294 UTC &     180 &      92.3 &   9.0 &       DIONE &  32 &        RHEA &  23 \\
ISS\_014TE\_MUTUALEVE004\_PRIME &2005 SEP 11 21:01:14.951 UTC &  100,1500 &      87.9 &  14.4 &      TETHYS &  33 &       DIONE &  33 \\
ISS\_015DI\_MUTUALEVE005\_PRIME &2005 SEP 16 08:15:40.484 UTC &    100,320,380.1800 &      65.7 &  12.6 &       DIONE &  56 &      TETHYS &  23 \\
ISS\_016EN\_MUTUALEVE002\_PRIME &2005 OCT 14 02:06:15.145 UTC &     150,260 &     108.5 &   6.8 &   ENCELADUS &  83 &      TETHYS &  23 \\
ISS\_016MI\_MUTUALEVE007\_PRIME &2005 OCT 14 19:36:15.742 UTC &     150 &      90.8 &  10.0 &       MIMAS &  15 &      TETHYS &   0 \\
ISS\_018TE\_MUTUALEVE001\_PRIME &2005 NOV 29 04:00:09.709 UTC &     150 &     122.2 &   6.6 &   TETHYS &   4 &      ENCELADUS &   4 \\
ISS\_018RH\_MUTUALEVE011\_PRIME &2005 DEC 05 03:46:40.402 UTC &     180 &     108.4 &  16.1 &       RHEA &  38 &        DIONE &  34 \\
ISS\_019EN\_MUTUALEVE008\_PRIME &2006 JAN 02 15:30:08.579 UTC &     150 &      97.0 &  15.8 &   ENCELADUS &  17 &       DIONE &  14 \\
ISS\_020RH\_MUTUALEVE001\_PRIME &2006 JAN 27 21:22:09.716 UTC &     180 &     123.2 &  18.1 &        RHEA &  57 &       MIMAS &  83 \\
ISS\_021RH\_MUTUALEVE002\_PRIME &2006 FEB 06 04:11:37.600 UTC &     180 &     110.0 &  25.8 &      RHEA &  30 &        TETHYS &  30 \\
ISS\_021MI\_MUTUALEVE004\_PRIME &2006 FEB 11 23:18:10.437 UTC &     150 &     101.3 &  22.5 &       MIMAS &  40 &      TETHYS &  40 \\
ISS\_021EN\_MUTUALEVE009\_PRIME &2006 MAR 02 12:09:40.192 UTC &     180 &     136.0 &  12.2 &   ENCELADUS &  39 &        RHEA &  23 \\
ISS\_021EN\_MUTUALEVE005\_PRIME &2006 MAR 03 03:20:09.844 UTC &     180,220 &     139.0 &  15.4 &   ENCELADUS &  35 &       DIONE &  35 \\
ISS\_022EN\_MUTUALEVE020\_PRIME &2006 MAR 16 08:42:10.643 UTC &     120 &      99.5 &  11.3 &   ENCELADUS &  30 &      TETHYS &  12 \\
ISS\_022RH\_MUTUALEVE009\_PRIME &2006 MAR 27 07:46:10.329 UTC &     680 &     158.7 &  13.9 &        RHEA &  10 &       DIONE &   8 \\
ISS\_023RH\_MUTUALEVE001\_PRIME &2006 APR 14 14:55:10.507 UTC &     180 &     130.8 &  20.5 &        RHEA &  10 &   ENCELADUS &  10 \\
ISS\_023DI\_MUTUALEVE006\_PRIME &2006 APR 17 05:29:10.052 UTC &     220 &     120.1 &  20.6 &       DIONE &  10 &        RHEA &  10 \\
ISS\_023RH\_MUTUALEVE006\_PRIME &2006 MAY 07 02:13:09.818 UTC &     820 &     159.8 &  16.1 &       RHEA &  10 &        DIONE &  10 \\
ISS\_024DI\_MUTUALEVE002\_PRIME &2006 MAY 14 06:43:40.157 UTC &     260,820,1000 &     133.8 &  15.9 &       DIONE &  31 &        RHEA &  31 \\
ISS\_024EN\_MUTUALEVE001\_PRIME &2006 JUN 06 16:01:40.167 UTC &     460 &     162.9 &  25.0 &      ENCELADUS &   0 &        TETHYS &   9 \\
ISS\_024EN\_MUTUALEVE009\_PRIME &2006 JUN 09 06:00:19.635 UTC &     680 &     161.2 &  23.3 &   ENCELADUS &  23 &        RHEA &   0 \\
ISS\_025MI\_MUTUALEVE001\_PRIME &2006 JUN 11 06:59:09.622 UTC &     460 &     158.5 &  23.7 &       MIMAS &   8 &   ENCELADUS &   9 \\
ISS\_025RH\_MUTUALEVE005\_PRIME &2006 JUN 11 08:30:39.537 UTC &     560 &     156.5 &  21.9 &        RHEA &  10 &      TETHYS &   8 \\
ISS\_025EN\_MUTUALEVE003\_PRIME &2006 JUN 13 15:00:10.227 UTC &     680 &     159.1 &  24.3 &   ENCELADUS &   8 &        RHEA &  10 \\
ISS\_025DI\_MUTUALEVE003\_PRIME &2006 JUN 14 03:38:09.998 UTC &     560 &     159.1 &  22.7 &       DIONE &  10 &      TETHYS &  10 \\
ISS\_025RH\_MUTUALEVE016\_PRIME &2006 JUN 16 10:46:09.733 UTC &     560 &     155.2 &  20.3 &        RHEA &   9 &      TETHYS &   0 \\
ISS\_025MI\_MUTUALEVE006\_PRIME &2006 JUN 21 22:58:09.851 UTC &     260 &     148.7 &  20.5 &      MIMAS &   2 &       TETHYS &   6 \\
ISS\_025MI\_MUTUALEVE007\_PRIME &2006 JUL 03 21:47:35.132 UTC &     460 &     145.6 &   9.8 &       MIMAS &   7 &       DIONE &  10 \\
ISS\_025RH\_MUTUALEVE004\_PRIME &2006 JUL 04 00:50:40.201 UTC &     180 &     141.8 &   8.1 &        RHEA &  10 &   ENCELADUS &   6 \\
ISS\_025RH\_MUTUALEVE006\_PRIME &2006 JUL 08 17:39:05.360 UTC &     680 &     158.4 &  13.4 &        RHEA &  20 &      TETHYS &  20 \\
ISS\_047TE\_MUTUALEVE004\_PRIME &2007 JUN 21 11:52:09.736 UTC &     120 &      64.2 &  12.3 &      TETHYS &  10 &   ENCELADUS &  10 \\
ISS\_084OT\_MUTGRNGAR001\_PRIME &2008 SEP 13 09:17:10.389 UTC &      60 &      32.7 &   5.3 &       DIONE &  27 &   ENCELADUS &  19 \\
ISS\_119MI\_MUTUALEVE001\_PRIME &2009 OCT 19 09:06:10.486 UTC &     150 &      94.9 &  12.1 &       MIMAS &  20 &        RHEA &  17 \\
ISS\_119RH\_MUTUALEVE001\_PRIME &2009 OCT 22 07:31:39.662 UTC &     180 &      98.0 &  12.6 &        RHEA &  20 &       DIONE &  18 \\
ISS\_120RH\_MUTUALEVE001\_PRIME &2009 OCT 26 20:22:09.847 UTC &     220 &     113.1 &  11.3 &        RHEA &  20 &      TETHYS &  17 \\
ISS\_120EN\_MUTUALEVE001\_PRIME &2009 OCT 27 01:47:09.742 UTC &     150 &     117.3 &  12.7 &   ENCELADUS &  20 &      TETHYS &  12 \\
ISS\_121DI\_MUTUALEVE001\_PRIME &2009 NOV 11 22:28:39.988 UTC &     260 &     118.8 &  14.8 &       DIONE &  20 &        RHEA &  18 \\
ISS\_121TE\_MUTUALEVE001\_PRIME &2009 NOV 11 23:25:09.968 UTC &     220 &     115.4 &  14.1 &      TETHYS &  20 &   ENCELADUS &  17 \\
ISS\_121EN\_MUTUALEVE001\_PRIME &2009 NOV 15 12:15:09.846 UTC &     120 &     115.2 &  13.7 &   ENCELADUS &  20 &        RHEA &  20 \\
ISS\_121RH\_MUTUALEVE002\_PRIME &2009 NOV 26 15:18:09.980 UTC &     180 &     103.6 &  10.5 &        RHEA &  20 &      TETHYS &  18 \\
ISS\_121DI\_MUTUALEVE002\_PRIME &2009 NOV 28 13:52:39.787 UTC &     180 &     109.5 &  13.1 &       DIONE &  20 &      TETHYS &  17 \\
ISS\_122RH\_MUTUALEVE003\_PRIME &2009 DEC 01 21:38:09.761 UTC &     150 &     107.3 &  12.3 &        RHEA &  20 &   ENCELADUS &  16 \\
ISS\_127DI\_MUTUALEVE002\_PRIME &2010 FEB 24 06:53:09.886 UTC &     150 &     116.6 &  12.2 &       DIONE &  22 &   ENCELADUS &  21 \\
ISS\_128DI\_MUTUALEVE002\_PRIME &2010 MAR 23 12:11:09.963 UTC &    460,680,2000 &      87.4 &   7.2 &       DIONE &  27 &       MIMAS &  16 \\
ISS\_128DI\_MUTUALEVE003\_PRIME &2010 MAR 26 19:03:09.955 UTC &    460,680,2000 &      89.9 &  11.7 &       DIONE &  27 &      TETHYS &  26 \\
ISS\_135DI\_MUTUALEVE001\_PRIME &2010 JUL 27 00:15:39.574 UTC &    150,460,560,2000 &      78.5 &   6.6 &       DIONE &  23 &        RHEA &  21 \\
ISS\_141DI\_MUTUALEVE002\_PRIME &2010 DEC 06 06:33:09.863 UTC &    320,460,1800 &      75.2 &  13.1 &      DIONE &  75 &       TETHYS &  75 \\
ISS\_143RH\_MUTUALEVE003\_PRIME &2011 JAN 20 18:18:09.919 UTC &    560,1000,2000 &      99.6 &  17.0 &       RHEA &  32 &        DIONE &  28 \\
ISS\_144RH\_MUTUALEVE002\_PRIME &2011 FEB 03 03:12:09.672 UTC &    460,560,2000 &      75.2 &   7.1 &        RHEA &  27 &       DIONE &  17 \\
ISS\_147RH\_MUTUALEVE006\_PRIME &2011 APR 25 19:25:40.172 UTC &    320,380,1500 &      66.9 &  13.3 &        RHEA &  30 &   ENCELADUS &  33 \\
ISS\_150DI\_MUTUALEVE006\_PRIME &2011 JUL 18 01:19:09.625 UTC &    560,680,2000 &      37.3 &  13.1 &       DIONE &  24 &        RHEA &  13 \\
ISS\_158RH\_MUTUALEVE001\_PRIME &2011 DEC 07 04:50:10.103 UTC &    460,680,2000 &      72.4 &  11.7 &       RHEA &  30 &        DIONE &  30 \\
ISS\_162TE\_MUTUALEVE002\_PRIME &2012 MAR 14 01:22:39.857 UTC &     150,180,820 &      51.7 &  10.1 &      TETHYS &  29 &       DIONE &  20 \\
\hline
\tablefoottext{a}{Mid-time for first image in sequence.}\\
\tablefoottext{b}{For primary satellite.}\\
\tablefoottext{c}{Zero indicates limb-fitting failed} 
\end{tabular}}
\end{table*}

Astrometric reduction was performed using the IDL-based $Caviar$ software package, developed at Queen Mary University of London, and incorporating the NAIF SPICE library \citep{Act96} together with the UCAC2 star catalogue \citep{Zach04}. Reduction consisted of a correction to the camera pointing direction for each image followed by an independent measurement of the centre-of-figure for each satellite using a limb-fitting approach. For this work, we used the Owen Model for the Cassini ISS NAC \citep{Owen03,Coop06} to relate the right ascension and declination of solar system objects and catalogue reference stars to their equivalent line and sample positions in each image. \Citet{Taj13} developed an alternative model, which may also be used. Unlike the Owen Model, the latter is more easily invertible, allowing line and sample to be more readily converted back to inertial positions, should that be required.

\subsection{Camera Pointing Correction}
For each image, the nominal camera pointing direction obtained from the SPICE `C-kernels' was corrected using an iterative minimisation of the observed-minus-computed positions of background reference stars, based on the UCAC2 catalogue. This involved firstly a manual translation of a template of predicted catalogue star positions, graphically, until approximately aligned with the imaged stars, followed by an automatic iterative search to fine-tune the alignment, typically with an accuracy of 0.1 pixel or less. The imaged star positions were computed using a centroiding technique based on the DAOPHOT method of \citet{Stet87}. A mean of 7.8 stars per image were detected, with a mean magnitude of 11.20$\pm$1.15.

\Citet{Taj13} showed how the number of detectable stars in a given image varied inversely with the size of the target satellite in the image: for very high resolution images, the satellite dominated the field-of-view, obscuring potential candidate stars for use in the pointing correction process. The mean resolution for the images used in the current work is 13.6 km/pixel. Based on the size of the largest satellite observed, Rhea, with a maximum radius of 764.30$\pm$1.10 km \citep{Thom07}, this corresponds to an imaged diameter of about 112 pixels at zero phase angle. Thus, unlike the observations used by \citet{Taj13}, which included some particularly high resolution images designed for the study of satellite surfaces, in this current work, the number of detected stars was less dependent on target size and more dependent on exposure length. Since the images used here were designed principally with astrometry in mind, exposure lengths were chosen, where possible, to balance the need to image often faint background stars against the requirements for optimum exposure of surface features.

\subsection{Satellite Position}
The pixel coordinates of the centre-of-figure of each satellite were estimated by comparing a shape model, projected onto the image, with the position of the imaged limb itself. Shape models, in the form of ellipsoids, were extracted from the latest Cassini SPICE kernels and projected on to the image using the chosen reference ephemeris (SAT360) and the corrected camera pointing information. Shape models were based on \citet{Thom07}.

Each imaged limb position was estimated by computing the maximum of the numerical derivative within a three-by-three array of pixels centred at a given pixel location. Detected limb points greater than two pixels away from the predicted limb position, based on the reference ellipsoid, were rejected. Different values were tested. Given that the camera pointing correction was applied before the limb-finding/fitting the measured limb positions were considered unlikely to be more than two pixels from their predicted locations based on the best available shape models. Using a progressively smaller value than 2.0 pixel reduced the scatter in the observed residuals (next section), but artificially drove the estimated limb points towards their predicted locations. It was considered more desirable to avoid this, at the expense of more random scatter in the residuals. 

An iterative fitting procedure was used to find the optimum alignment of the position of the limb (based on the shape model) with the imaged limb positions. The observed position of the centre-of-figure was then computed by correcting the predicted position by the mean shift required to align, optimally, the measured and predicted limbs. Five ellipse parameters could potentially be fitted. However, the ellipsoid projection fixes the size of the ellipsoid, while its orientation is fixed by the satellite's known orientation to the order of 0.1 degree \citep{Arch11} and the camera's twist angle (known to order of 60 $\mu$rad \citep{Porc04} and 90 $\mu$rad \citep{Taj13}, leaving the ellipse's centre to be fitted only.

The measured pixel coordinates (line versus sample) for each satellite are shown in Fig. 2.

\begin{figure}
\resizebox{\hsize}{!}{\includegraphics{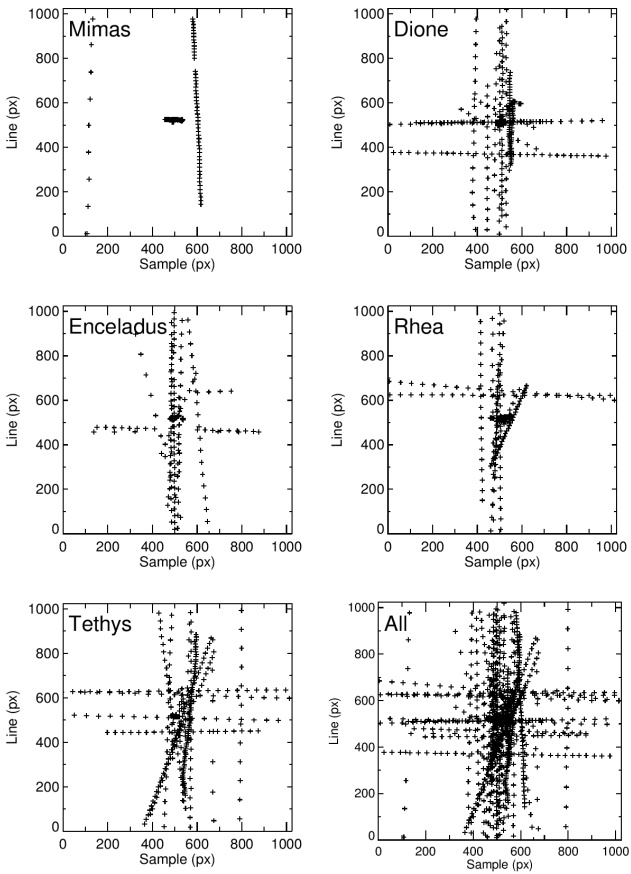}}
\label{Fig. 2}
\caption{Observed image positions of satellite centres, line versus sample. Image sizes are 1024 by 1024 pixel. }
\end{figure}

\subsection{Reduced Data}
To allow more flexibility in the use of the reduced data, we provide separate positions for each satellite, rather than inter-satellite coordinates, particularly given that some images ($\sim$15$\%$ of the total) contain only one satellite. 

\begin{table*}
\caption{Sample of Cassini ISS Observations\label{tbl2}}
\label{table:2}
\centering
\resizebox{\textwidth}{!} {
\begin{tabular}{ccrrrrrrrr}
\hline\hline
Image ID &Mid-time (UTC) &$\alpha_{c}$ &$\delta_{c}$ &TWIST &Line\tablefootmark{a} &Sample\tablefootmark{a} &$\alpha$ &$\delta$ &Body\\
&&(deg)&(deg)&(deg)&(px)&(px)&(deg)&(deg)&\\
\hline
N1487595425 &2005 FEB 20 12:30:38.277 & 27.296799 & -6.438903 &178.585994 &  595.83 &  591.92 & 27.268304 & -6.467161 &DIONE       \\
N1487595459 &2005 FEB 20 12:31:12.277 & 27.289024 & -6.438771 &178.630998 &  595.78 &  591.47 & 27.260708 & -6.467036 &DIONE       \\
N1487595459 &2005 FEB 20 12:31:12.277 & 27.289024 & -6.438771 &178.630998 &  621.95 &  994.47 & 27.121289 & -6.472685 &RHEA        \\
N1487595493 &2005 FEB 20 12:31:46.276 & 27.281160 & -6.438414 &178.595788 &  596.02 &  590.78 & 27.253062 & -6.466749 &DIONE       \\
N1487595493 &2005 FEB 20 12:31:46.276 & 27.281160 & -6.438414 &178.595788 &  622.42 &  947.46 & 27.129638 & -6.472789 &RHEA        \\
\hline
\end{tabular}}
\tablefoot{Columns $\alpha_{c}$, $\delta_{c}$ and TWIST refer to the right ascension, declination and twist angle of the camera's pointing vector in the International Celestial Reference Frame (ICRF), while $\alpha$ and $\delta$ are the right ascension and declination in the ICRF for the body listed in the far right-hand column. Relative to the SAT360 ephemeris, considering all the individual positions together, we obtain a precision of 0.34 pixels in line and 0.41 in sample. The full table is available from the CDS.
\tablefoottext{a}{The origin of the line, sample coordinate system is at the top left of the image with line, y, increasing downwards and sample, x, to the right. Image size is 1024 by 1024 pixels.}}
\end{table*}

Also, following \citet{Coop06,Taj13}, in addition to providing the measured satellite positions in right ascension and declination ($\alpha$, $\delta$) in the International Celestial Reference Frame (ICRF), we provide their measured pixel coordinates and camera pointing directions, so that either of these sets of measurement can easily be re-estimated independently at some future date, if required. This would, for example, allow the camera pointing corrections to be updated using improved star catalogues, such as those soon to become available from the GAIA mission, or for the astrometric positions themselves to be re-estimated independently of the pointing corrections, if advances in limb measurement or centre-finding became available (see also the section on Sources of Error). Alternatively, a different NAC camera distortion model, such as that developed by \citet{Taj13}, could also be used.

The complete set of reduced data is available at the CDS. A small section of the table showing the satellite positions is reproduced in Table 2. Computed star positions (Table 5) are only available electronically at the CDS.

\section{Analysis of Residuals}
\noindent
In this work, we compared observed minus computed (O$-$C) residuals relative to two different ephemerides: JPL's SAT360 ephemeris, based on a fit to Earth-based, Pioneer, Voyager, HST and Cassini data to the end of 2013, and NOE-6-2012-MAIN, created by IMCCE Paris based on a fit to Earth-based and spacecraft data from 1886-2012. In both cases, these are post-fit residuals, since the orbit models on which both these ephemerides are based included the observations presented here. However, SAT360 was generated based on corrected astrometry, which included a constant sample and line bias due to camera pointing plus additional corrections for the satellite-dependent phase biases (R.A. Jacobson, private communication).

In Fig. 3, we show O$-$C residuals relative to the SAT360 ephemeris for the 976 pairs of observations. Fig. 3(a) shows the residuals, line versus sample, for each individual observation, computed using the absolute satellite positions, while Fig. 3(b) shows the equivalent residuals based on the separation between the primary and secondary satellite in each image. The mean line and sample residuals for the absolute positions in Fig. 3(a) are 0.27 and $-$0.11 pixel respectively, with $\sigma$ values of 0.34 and 0.37 pixel, while for the inter-satellite separations (Fig. 3(b)), we obtain an order-of-magnitude improvement in accuracy, with means of -0.03 and $-$0.01 pixel respectively and $\sigma$ values of  0.29 and 0.26 respectively. The equivalent mean residual values across all absolute positions (2303 values) in km are 4.3 km (line) and $-$2.4 km (sample), with $\sigma$ values of 5.6 and 7.0 km, respectively. By comparison, \citet{French06} obtained typical uncertainties at Saturn of 80km and 120km respectively, using the Planetary Camera and the Wide Field Planetary Camera 2 of the Hubble Space Telescope.

We discuss these results further in the following section. 

\begin{figure}
\resizebox{\hsize}{!}{\includegraphics{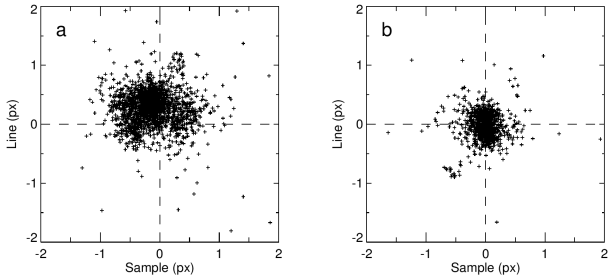}}
\label{Fig. 3}
\caption{Comparison between O$-$C residuals based on absolute positions and those based on inter-satellite differences, plotted as line residual versus sample residual relative to the JPL SAT360 ephemeris. All satellites are shown. Units are NAC pixels.}
\end{figure}

Line and sample residuals for each individual satellite (absolute positions) are plotted as a function of time relative to the SAT360 ephemeris in Fig. 4 and as line versus sample in Fig. 5. 

\begin{figure}
\resizebox{\hsize}{!}{\includegraphics{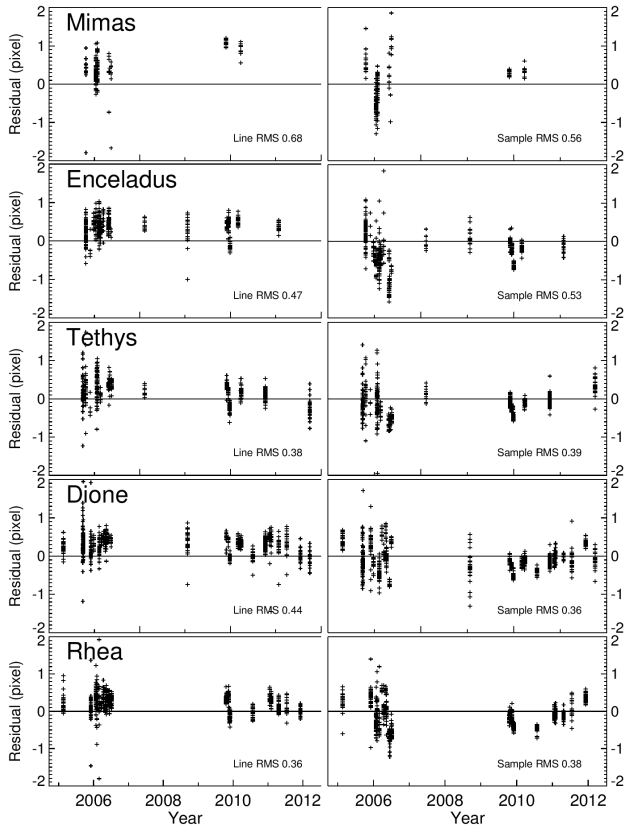}}
\label{Fig. 4}
\caption{O$-$C residuals for each satellite, using absolute positions, plotted versus time relative to the JPL SAT360 ephemeris. For each satellite, line residuals are plotted on the left, and sample on the right. Units are NAC pixels.}
\end{figure}

\begin{figure}
\resizebox{\hsize}{!}{\includegraphics{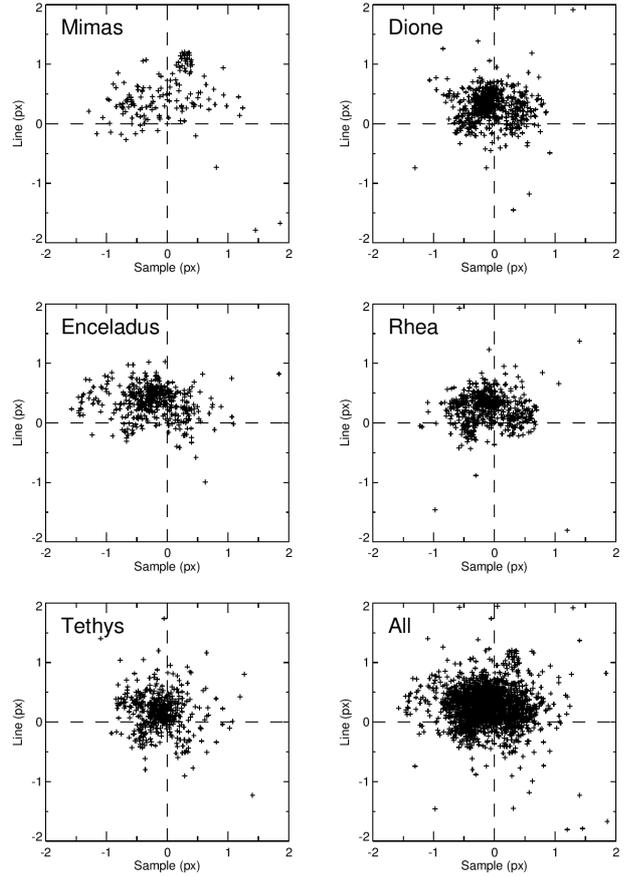}}
\label{Fig. 5}
\caption{O$-$C residuals for each satellite, using absolute positions, relative to the JPL SAT360 ephemeris, plotted as line residual versus sample residual. Units are NAC pixels.}
\end{figure}

\section{Sources of Error}

\Citet{Taj13} give a detailed description of the sources of uncertainty in the astrometric reduction of Cassini ISS data. Here we focus on possible systematic errors in the absolute positions, that might give rise to the non-zero means of several tenths of a pixel, mentioned in the previous section. Although these biases are small in comparison to the typical accuracy achieved with earth-based astrometry \citep{French06}, and we have seen in the previous section that they can be largely eliminated by measuring inter-satellite separations, it is clearly desirable to understand their origin.

\begin{table}
\caption{Mean values of residuals in pixels relative to the JPL SAT360 ephemeris, including standard deviations.}
\label{table:3}
\centering
\begin{tabular}{lcccc}
\hline\hline
&  line & $\sigma_{line}$ & sample & $\sigma_{sample}$\\
\hline
Mimas & 0.44 &0.52& -0.04&0.56\\ 
Enceladus & 0.35 &0.32&-0.24&0.47\\
Tethys & 0.16 &0.35&-0.13&0.36\\
Dione &0.30  &0.32& -0.07&0.36\\
Rhea & 0.24 &0.28& -0.11&0.37\\
\hline
All & 0.28 &0.34 & -0.12&0.41\\
\hline
All (inter)\tablefootmark{a}& -0.03 &0.29 & -0.01&0.26\\
\hline
\end{tabular}
\tablefoot{
\tablefoottext{a}{Using inter-satellite separations between pairs of satellites within a given image.}
}
\end{table} 

\begin{table}
\caption{Mean values of residuals in pixels relative to the IMCCE NOE--6-2012-MAIN ephemeris, including standard deviations.}
\label{table:4}
\centering
\begin{tabular}{lcccc}
\hline\hline
&  line & $\sigma_{line}$ & sample & $\sigma_{sample}$\\
\hline
Mimas & 0.55 &0.57& -0.25&1.19\\ 
Enceladus & 0.32 &0.42&-0.30&0.40\\
Tethys & 0.18 &0.47& 0.01&0.52\\
Dione &0.19  &0.38& -0.04&0.35\\
Rhea & 0.26 &0.31& -0.08&0.39\\
\hline
All & 0.25 &0.42 & -0.10&0.51\\
\hline
All (inter)\tablefootmark{a}&0.04 &0.41 & -0.08&0.30\\
\hline
\end{tabular}
\tablefoot{
\tablefoottext{a}{Using inter-satellite separations between pairs of satellites within a given image.}
}
\end{table} 

As noted previously, Figs. 3-5 indicate a non-zero mean of 0.28 pixels in the line residuals across all five satellites. This is also clear quantitatively in Tables 3 and 4 where we show mean values relative to two different ephemerides. The mean line residuals are consistently positive, although there is some variation in the magnitude of the mean, both between satellites, and also depending on the reference ephemeris used. The largest positive mean values occur in the line residuals for Mimas and Enceladus, based on either of the reference ephemerides (SAT360 and NOE-6-2012-MAIN). Given that these are post-fit residuals, the implication is either that there is a missing or poorly-determined dynamical component in the models used to generate both ephemerides, or that there is a systematic bias in the measurements, or a combination of both. Since all five satellites show a positive mean at some level, an inadequacy in the modelling would have to affect all five satellites in the same direction, which seems less likely than a systematic bias in the measurements. Also, any inadequately modelled component in the dynamics would give rise to a bias in both absolute and relative positions, which is not observed.

Some measurement bias is inevitable because the phase angle is never precisely zero in practice, so that imaged limbs are always one-sided, with the terminator forming what remains of the boundary of the satellite image. This is true for all resolved observations based on limb measurement. \citet{Taj13} also found a positive bias in the residuals for Mimas and Enceladus in the direction towards the Sun, with a significantly larger bias for Mimas than Enceladus. They put forward two possible explanations: (1) that the greater level of cratering on Mimas may distort the limb and (2) that the dimensions of Mimas could be larger than those based on the shape model of \citet{Thom07}. However, in the data presented here, this effect is systematic across all the observations for five different satellites, implying that some common geometric characteristic of the dataset as a whole is the key contributory factor: if the limbs had randomly distributed phases and illumination directions, this effect would not be systematic. A clue that such a common geometric characteristic exists is evident from Fig. 2, showing how the measurements are distributed dominantly along the line axis.

We investigated this further using synthetically-generated images, whose centre-of-figure is known in advance. Sequences of images were generated, using the $Mathematica$ software package \citep{Math12} for different illumination directions and phase angles, in order to assess possible measurement bias as a function of these parameters. In Figs. 6 and 7, we show using these synthetic images how the limb-fitting algorithm does indeed generate a bias in the computed centre-of-figure that varies as a function of phase angle and illumination direction. The maximum size of the observed bias is more than 1.5 pixel, which is almost an order of magnitude larger than the mean values we obtain for the real images. This discrepancy arises because the illumination model used to generate the synthetic images is not a realistic photometric match for the real images. Thus this comparison is illustrative only and serves purely to demonstrate how a bias can arise and how it may change according to the imaging geometry.

Returning to the question of why this appears as a systematic effect in the real data, in Fig. 8(a), we plot the variation of phase angle across the entire set of real images. This shows clearly that phase angles across all the observations are clustered around 90--100 degrees (see also the mean phase values listed in Table 1). Furthermore, Fig. 8(b) shows that the sun directions are also preferentially distributed along the positive line direction. We conclude therefore that the combination of these two geometrical characteristics of the images accounts for the systematically positive mean values in the absolute positions: from the synthetic tests, the maximum possible bias occurs at $\sim$90 degree phase angle, and the common alignment of the limbs in the real images in the positive line direction then accounts for the appearance of a systematic positive bias in that direction across all the absolute positions.

\begin{figure}
\resizebox{\hsize}{!}{\includegraphics{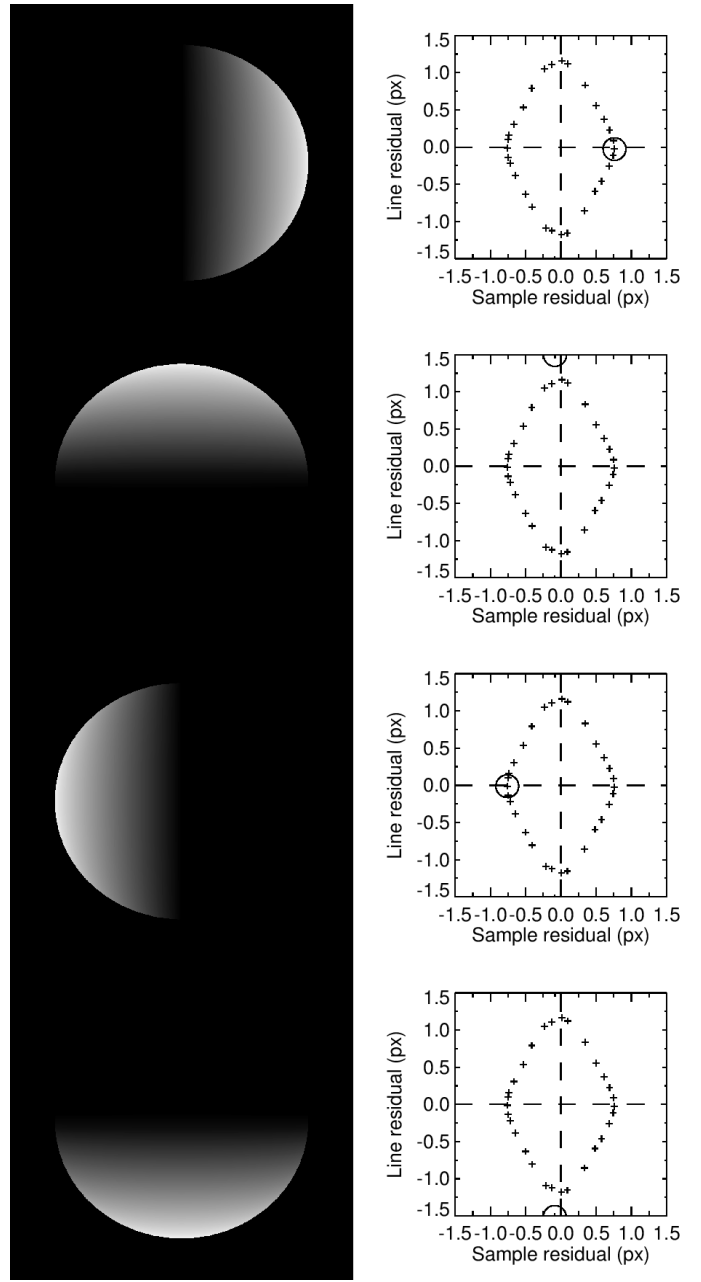}}
\label{Fig. 6}
\caption{Synthetic images for different illumination directions, showing O$-$C residuals in pixel (right) for centre-of figure 
positions measured by limb-fitting the images shown left. The bias is always in the illumination direction. All points in the right-hand displays 
are the same, with the particular O$-$C residual values corresponding to each of the four images represented by the small circles (directions
are, from top to bottom, 0 deg, 90 deg, 180 deg, 270 deg). }
\end{figure}

\begin{figure}
\resizebox{\hsize}{!}{\includegraphics{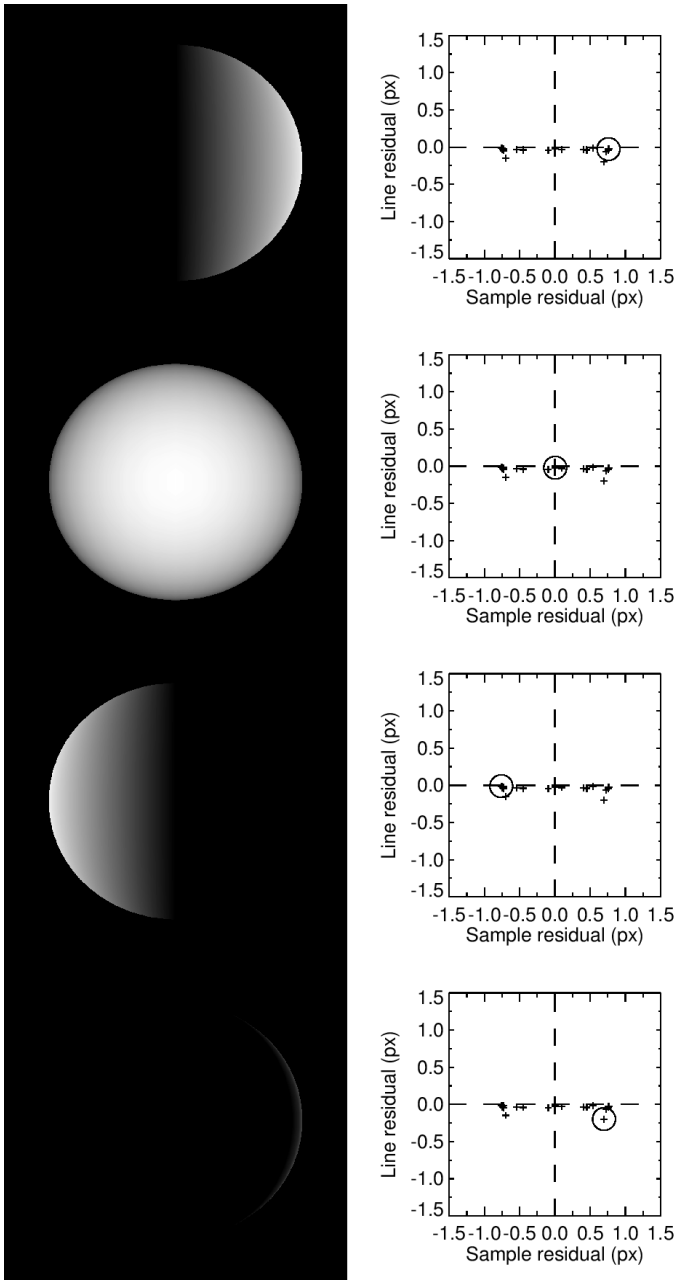}}
\label{Fig. 7}
\caption{Synthetic images for different phase angles, showing O$-$C residuals in pixel (right) for centre-of figure 
positions measured by limb-fitting the images shown left. The maximum bias occurs for 90 degree phase and is zero for zero
phase. All points in the right-hand displays are the same, with the particular O$-$C residual values corresponding to 
each of the four images represented by the small circles (phase angles are, from top to bottom, +90 deg, 0 deg, -90 deg,
180 deg).}
\end{figure}

\begin{figure}
\resizebox{\hsize}{!}{\includegraphics{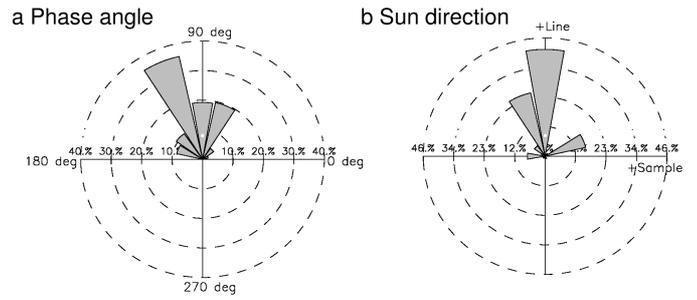}}
\label{Fig. 8}
\caption{Distribution of (a) satellite phase angles and (b) sun directions for the Cassini images used in this work. Sun directions are resolved into line and sample directions within each image.}
\end{figure}

\section{Discussion and Conclusions}

A particular advantage of all mutual event observations, using either ground-based or spacecraft imaging, is that image-dependent errors cancel when computing the separation between two satellites in the same image, leading to a potentially significant increase in accuracy. This type of observation is also useful when no suitable reference stars are detectable within a given image. Although we did not foresee the tendency of the limb-fitting method to introduce a bias in the absolute positions of satellite centres-of-figure in the direction of the limb (i.e. in the sun direction), we have shown that this also cancels when computing inter-satellite separations because the magnitude and direction of the the bias turns out to be dependent on phase angle and sun direction, and these two quantities, like the camera pointing direction, are also effectively constant for a given image. 

On the other hand, a single measurement of the separation between two satellites necessarily contains less information than the individual absolute positions combined. Thus there is an inevitable trade-off in information content versus accuracy, depending on which of the two approaches to data reduction is used. This is one reason why we have chosen to provide the raw data in terms of the absolute positions, in order to give the reader the option to use either approach.

As described previously, investigation showed that Sun orientations (and phase angles) are preferentially distributed in one direction across the dataset as a whole and thus the bias, rather than being randomly distributed, has a preferred orientation in that direction (the positive line direction). This is an unintended consequence of the particular observing geometry of the image sequences, Cassini Project standard operational practice and the habits of the imaging sequence designers.

Given that there may be circumstances where it is more desirable to make use of the greater information content of the absolute positions, the question arises as to what can be done either to correct such biases, or to minimise their occurrence in the first instance. We have considered four approaches: (a) correcting an existing bias using information derived from synthetic images, (b) correcting an existing bias by subtracting mean values derived for each satellite, (c) changing the observation strategy to prevent the bias from occurring systematically in one direction and (d) improving the performance of the limb-measuring technique to try to reduce the magnitude of any bias.

Firstly, considering (a), the synthetic images have provided a useful way of evaluating how the measured centre-of-figure based on limb-fitting behaves as a function of phase angle and illumination direction. Thus, in principle, a table of corrections could be generated from these synthetic images, corrections which could then be applied to the real observations in order to remove any potential bias generated in the limb-fitting. However, as noted previously, the magnitude of the bias we currently obtain from the synthetic images is much larger than any effect we see in the real observations, so this approach would only be viable if a way could be found to model the limb profiles to give a satisfactory photometric representation of the profiles obtained from the real images. Although a photometric approach is routinely used for ground-based observations of mutual phenomena for unresolved bodies (see for example \citet{Noy03,Arlot08}), to our knowledge this has not been attempted for observations of resolved satellites, where centre-of-figure measurements are based on limb-fitting. Clearly, with an observed mean bias of less than 0.2 pixel for some satellites, any such correction process would have to be precise enough so as not to risk introducing a bias in a different direction. This will be the subject of further work.

In principle, biases in the absolute positions could also be corrected by applying a global correction for each satellite, prior to orbit modelling, based on the computed mean values of the O$-$C residuals relative to a suitable reference ephemeris. However, clearly this can only be justified if there is sufficient confidence that the mean values do not represent genuine dynamical effects.

A different approach to observing the mutual events with the aim of randomising the sun directions would potentially reduce the tendency of the bias to be systematic in one direction. However this would only randomise the direction of the bias across the dataset as a whole and would not reduce its magnitude in a given image. Also any such modification of the observing strategy would still need to satisfy spacecraft flight rules.

A possible fourth approach would be to improve the performance of the limb-measurement algorithms in order to minimise the magnitude of any possible bias in a given image. Approaches which we have investigated so far have included the application of a gaussian to the derivative in order to estimate the limb position to sub-pixel precision, the use of other edge detection algorithms, such as the Canny algorithm \citep{Canny86}, and the interpolation of the input images to produce finer spatial sampling, before limb fitting. Preliminary investigations have shown that all these approaches, to a greater or lesser degree, can provide some improvement in the overall precision of the data, but do not significantly increase the accuracy. 

Although these are small effects, which we already have seen can be largely eliminated by measuring the separations between pairs of satellites in each mutual event image, further investigation would still be advantageous, particularly given the need for ever greater astrometric accuracy based on current scientific goals.

\begin{acknowledgements}
The authors thank the reviewer, Dr Marina Brozovic, for helping us to improve this paper. NJC is grateful to the Paris Observatory for funding while he was an invited researcher at the IMCCE. The work was also supported by the Science and Technology Facilities Council (Grant No. ST/F007566/1) and NJC and CDM are grateful to them for financial assistance. CDM is also grateful to The Leverhulme Trust for the award of a Research Fellowship. VL and RT thank UPMC for funding under grant EME 0911. NJC, CDM, VL and RT also thank the CNRS and Royal Society for funding under the `GAME' proposal and the ESPACE consortium for funding under agreement 263466. The authors thank the members and associates of the $Cassini$ ISS team, particularly Kevin Beurle for assistance in designing some of the image sequences used in this work, and also thank their colleagues in the Encelade working group at the IMCCE of the Paris Observatory (http://www.imcce.fr/$\sim$lainey/Encelade.htm) and Dr Bob Jacobson at JPL for many discussions. 
\end{acknowledgements}

\end{document}